# Coherent modulation of the electron temperature and electron-phonon couplings in a 2D material

Yingchao Zhang[1]†, Xun Shi[1]*†, Wenjing You[1]†, Zhensheng Tao[1,2], Yigui Zhong[1], Fairoja Cheenicode Kabeer[3], Pablo Maldonado[3], Peter M. Oppeneer[3], Michael Bauer[4], Kai Rossnagel[4,5,6], Henry Kapteyn[1], Margaret Murnane[1]

[1]Department of Physics and JILA, University of Colorado and NIST, Boulder, CO 80309, USA

[2]State Key Laboratory of Surface Physics and Department of Physics, Fudan University, Shanghai, China

[3]Department of Physics and Astronomy, Uppsala University, Box 516, S-75120 Uppsala, Sweden

[4]Institute of Experimental and Applied Physics, Kiel University, D-24098 Kiel, Germany

[5]Deutsches Elektronen-Synchrotron DESY, D-22607 Hamburg, Germany

[6]Ruprecht Haensel Laboratory, Kiel University and DESY, D-24098 Kiel and D-22607 Hamburg, Germany

*Corresponding author. Email: Xun.Shi@colorado.edu

†These authors contributed equally to this work.

Ultrashort light pulses can selectively excite charges, spins and phonons in materials, providing a powerful approach for manipulating their properties. Here we use femtosecond laser pulses to coherently manipulate the electron and phonon distributions, and their couplings, in the charge density wave (CDW) material 1$T$-TaSe$_2$. After exciting the material with a short light pulse, spatial smearing of the electrons launches a coherent lattice breathing mode, which in turn modulates the electron temperature. This indicates a bi-directional energy exchange between the electrons and the strongly-coupled phonons. By tuning the laser excitation fluence, we can control the magnitude of the electron temperature modulation, from ~ 200 K in the case of weak excitation, to ~ 1000 K for strong laser excitation. This is accompanied by a switching of the dominant mechanism from anharmonic phonon-phonon coupling to coherent electron-phonon coupling, as manifested by a phase change of π in the electron temperature modulation. Our approach thus opens up possibilities for coherently manipulating the interactions and properties of quasi-2D and other quantum materials using light.

The interactions between the charge and lattice degrees of freedom in materials play a pivotal role in determining the state and properties of matter [1–3]. For example, electron-phonon and electron-electron interactions are believed to be key to stabilizing the periodic lattice distortion (PLD) and corresponding charge density modulation in many CDW materials [1]. Understanding and coherently manipulating these strong interactions is a grand challenge in quantum materials. Under thermal equilibrium conditions, the phase diagram of a material can be tuned by varying the temperature, pressure, number of layers, or chemical doping. However, some "hidden" states of a material [4,5] are simply unreachable via equilibrium tuning [1,2].

Additional pathways for accessing new couplings and states can open up by driving the material out of equilibrium. For example, mid-infrared or terahertz pulses can resonantly excite selected phonons in a material, to induce interesting new phenomena such as enhanced superconductivity [6–9], hidden states [10] and phonon upconversion [11]. Such sophisticated excitations require precise knowledge of the phonon spectrum and fine tuning of the THz excitation field. Another widely-used approach takes advantage of near-infrared laser pulses to first excite electrons, which later relax by coupling to the phonon bath [12,13]. In equilibrium, phonons in a material will generally have random phases. However, an ultrashort laser pulse can excite coherent phonons with well-defined phase relations, via impulsive stimulated Raman scattering (ISRS) or displacive excitation (DECP) mechanisms [14]. Experimentally, coherent phonons can be detected using transient reflectivity [14], diffraction [15,16] or photoemission [16–19]. The excitation and relaxation of such coherent modes relies on the symmetry and electronic properties and provides a unique opportunity to study the mode-projected electron-phonon and phonon-phonon couplings [14,20,21]. Indeed, the coherent amplitude modes in quasi-2D CDW materials such as $1T$-TaS$_2$ and $1T$-TaSe$_2$ have been extensively studied [17–19,22–24]. The previous measurements have observed the modulation of the band position [17,18] or chemical potential [25] after exciting the material with a femtosecond laser. However, to date the nature of the strong coupling between the coherent phonon mode and the electrons, which underlies the CDW phase itself, remains difficult to probe.

Here we use ultrafast light pulses to drive the quasi-2D CDW material 1$T$-TaSe$_2$ into a nonequilibrium state, thereby allowing us to coherently manipulate the electron and phonon distributions and couplings. Femtosecond laser excitation will very rapidly smear the spatial charge modulation which launches a coherent breathing mode within ~100fs. We then use time- and angle-resolved photoemission spectroscopy (trARPES) [26,27] and ultrafast electron calorimetry [19] to measure the dynamic electron temperature, which allows us to make two new observations. First, we observe a large coherent modulation of the electron temperature (between ~200 and ~1000 K) that is superimposed on the monotonic relaxation of the hot electrons — indicating a bi-directional exchange of energy between the electron and strongly-coupled phonon bath. Second and most interestingly, this electron temperature modulation is in phase with the coherent phonon-driven band oscillation at low fluence, but exhibits a phase change of π at a critical laser fluence that launches a new long-lived metastable CDW state. Our simulations show that this π phase change is associated with a switching of the dominant cooling mechanism from anharmonic phonon decay to coherent electron-phonon coupling, as illustrated in Fig. 1, showing that by tuning the ultrafast laser excitation, we can control the electron-phonon interactions. This approach can also uncover new routes for light-assisted control of interactions in other strongly-coupled materials including superconductors and semiconductors.

As a prototypical CDW material, 1$T$-TaSe$_2$ is a layered dichalcogenide consisting of hexagonal Ta and Se layers. In the CDW state below $T_c$ of 470 K, the lattice reconstructs into a $\sqrt{13} \times \sqrt{13}$ star-of-David supercell [28]. The corresponding spectroscopic signatures are electronic band folding and the appearance of an energy gap [29–31] as shown in the data in the left panel of Fig. 2. Upon excitation by a 1.6 eV ultrafast laser pulse, the CDW amplitude mode (the breathing mode of the stars) can be coherently excited, as illustrated in Fig. 1. This coherent phonon mode, with a frequency of about 2 THz, is immediately observed in the oscillation of the Ta 5$d$ band as plotted in the right panel of Fig. 2, consistent with previous reports on a similar material [17,18].

In addition to the oscillatory binding energy of the Ta 5$d$ electronic band, which determines the available states, we also investigate the transient electron temperature — which is related to the electron occupation near the Fermi level ($E_F$). Both the band position and the electron temperature are simultaneously extracted from the trARPES spectra, by globally fitting the energy distribution curves (EDCs) at each time delay with a Lorentzian function multiplied by the Fermi-Dirac distribution. In Fig. 3a we plot the electron temperature together with the band shift dynamics at a relatively low fluence of 0.24 mJ/cm$^2$, where oscillatory components at ~ 2 THz are clearly observed along with the expected relaxation of the electron temperature. This oscillation has the same frequency as that of the band shift, and thus originates from the coupling of the electron bath to the coherent CDW amplitude mode. Moreover, the electron temperature and the band shift oscillate in phase. However, for fluences higher than a critical value of 0.7 mJ/cm$^2$ (Fig. 3b), the oscillations are out of phase and strongly damped. The instantaneous decay rates determined as $g(t) = -\frac{d}{dt}\ln[T_e(t)]$, from the electron temperature $T_e$ in Fig. 3a and b are shown in Fig. 3c and d, respectively, to better highlight the oscillatory nature of the electron temperature (see Supplementary Information).

These findings are novel in the view of decades of past research where the thermalized electron temperature after ultrafast laser excitation was always observed to decay monotonically due to electron phonon scattering causing an energy flow from hot electrons to the lattice [12]. *In contrast, in our data, the electron temperature can increase significantly by up to 1000 K (Fig. 3a and b), and the decay rate can be negative (Fig. 3c),* indicating an extremely strong coherent coupling between the electrons and the CDW amplitude mode (see Supplementary Information). Clearly our data show that in the presence of the excited coherent phonon, the overall electron-phonon coupling cannot be treated as a single constant averaged in the time domain. Instead, the coherent phonon modulates the electron relaxation process, effectively compressing and expanding the electron occupation (and likely the density of states) near $E_F$ to give rise to a modulation of the electron temperature.

We also plot the instantaneous decay rate of the band shift for comparison (Fig. 3c and d), from which a change of π in the phase difference from low to high fluence is confirmed. Note that this flip of oscillation further validates the reliability of our data by excluding the possible artifacts from data fitting. In order to systematically investigate the phase relation between the electron temperature and band shift, we model the dynamics as a classical damped harmonic oscillator [14], as shown in Fig. 3a and b (see Supplementary Information). We plot in Fig. 4a the extracted oscillation phase $\phi$ of the electron temperature and the band shift, they are similar at low fluences and deviate from each other at high fluences. The phase difference in Fig. 4b clearly indicates the π phase change at a critical fluence $F_c$. Note that there is no data point near $F_c$ for the phase of the electron temperature (red circle in Fig. 4a), because there is no observable oscillation at this fluence within our experimental uncertainties (see Supplementary Information). This is consistent with the oscillation phase change at $F_c$ — two oscillations with a phase difference of π would suppress or cancel each other.

It is worth mentioning that this phase change in the electron temperature oscillation occurs at the critical fluence at which the transient metastable state appears (Fig. 4c). This is accompanied by an abrupt reduction of the coupling of electrons to some phonon modes while at the same time an increased coupling to others (Fig. 4d) [19]. This demonstrates that the observation of any change in phase of the $T_e$ oscillation is also a sensitive fingerprint of a transient phase transition. To date, the measurements of coherent phonons, such as their frequency and damping constant, were observed to change only during an equilibrium phase transition involving lattice reorganizations [22].

Next, we discuss how the electron temperature could be modulated coherently by the CDW breathing mode, and also what might give rise to the change in phase of π observed between the $T_e$ modulation and band shift as the laser fluence is increased. In previous work we showed that above a critical laser fluence that corresponds to 1$T$-TaSe$_2$ entering a new metastable CDW state, the electron-phonon coupling to some phonon modes is enhanced selectively, while the couplings to others are reduced [19]. In the case of laser excited quasi-2D materials, the timescales for subsequently exciting the different phonon baths can be different [19,32]. It means that ultrafast laser excitation is a unique way for isolating the interactions between the charge and the strongly coupled phonons (often in-plane), since weakly-coupled phonons (often cross-plane) will not be excited until greater than 100 picoseconds. This implies that there could be dramatic variations in several electron-phonon interactions. The π phase change suggests that the electron bath is *coherently* coupled to more than one phonon bath, and the relative coupling strength of each channel varies between the normal and new metastable CDW states. Indeed, the coherent CDW breathing mode can be very strongly excited by laser-induced charge smearing, reaching lattice distortions not possible under equilibrium conditions [19,33]. Thus, it is not surprising that its coupling to electrons and other phonon modes (e.g. anharmonic decay to low energy phonon modes) should be considerable. The latter was observed in the coherent $A_{1g}$ phonon of photoexcited bismuth by x-ray diffuse scattering [34].

Based on these insights into the normal and new metastable CDW states in 1$T$-TaSe$_2$, we describe the electron relaxation process after the pump as (see Supplementary Information),

$$C_e \frac{\partial T_e}{\partial t} = -G_0(T_e - T_l) - G_{An}(T_e - T_{An}) - f\frac{\partial Q}{\partial t} \qquad (1)$$

The first term is the widely considered incoherent electron-phonon scattering, that accounts for unidirectional heat transfer from the electrons to the lattice [12]. A recent theoretical study suggested that the presence of several phonon modes ν**k** with mode-dependent couplings can alter the direction of energy flow between phonons and electrons [35]. Past work in graphene suggested conversely that a THz-driven vibrational mode can induce a dynamic renormalization of the electronic density of states, the phonon spectrum and thus the electron-phonon coupling [36,37]. In the case of 1*T*-TaSe$_2$ studied here, the strongly-coupled breathing mode drives the atomic displacements periodically over large lattice distortions. It is therefore reasonable to consider that a variation of overall electron-phonon coupling $G_0$ could give rise to the $T_e$ oscillation [38]. However, the phase of such an oscillation would have a π/2 difference from that of the band shift. This mismatch rules out the possibility that a modulation of $G_0$ is the main origin of the electron temperature oscillation.

The second term represents the coupling of the electrons to the low energy phonon modes via anharmonic decay of the coherent breathing mode. This anharmonic coupling modulates the effective temperature $T_{An}$ of the low energy modes with the same frequency as the breathing mode, but with a phase difference of π/2 [34,39]. Equation (1) predicts that this will result in an electron temperature $T_e$ oscillation that is in phase with the band shift, consistent with our data at low laser fluences. As shown in Fig. 3e, Eq. (1) reproduces all the fine details of the data shown in Fig. 3a, when a dominant contribution from this anharmonic decay process is included (see Supplementary Information).

The third term describes the coherent coupling between the electrons and the amplitude mode, where $f = M\omega_0^2 Q_0$ is the force, $M$ is the mode's mass, $\omega_0$ is the angular frequency, $Q_0$ is the quasi-equilibrium coordinate, $Q$ is the mode's coordinate. The coherent amplitude mode is initially excited by charge smearing, which in turn modulates $T_e$ through coherent electron-phonon coupling. In contrast to the second term, it modulates $T_e$ directly — and with a π phase shift from the oscillations of the band shift [40]. This explains the origin of the π phase shift at fluences above the critical fluence $F_c$ required to access the new CDW metastable state. As shown in Fig. 3f, we can reproduce all aspects of the experimental data from Fig. 3b by including a dominant contribution from this coherent electron-phonon coupling.

Our simulations thus reveal a competition between multiple couplings that change in relative dominance as the material enters a new CDW state. At fluences above $F_c$, our fit shows that the electron coupling to low energy phonon modes dramatically weakens — the dominant contribution switches from an anharmonic phonon-phonon coupling to a coherent electron-phonon coupling (see Supplementary Information). This switch accompanies an ultrafast phase transition to a CDW metastable state, where the overall electron-phonon coupling changes from homogeneous to mode-selective [19]. In the future, it will be interesting to probe the dynamic phonon spectrum as it evolves over fs to ns timescales, as well as to directly image the coupled charge lattice motions using a combination of ultrafast X-ray, electron and Raman techniques.

From a physical point of view, the electron-phonon and phonon-phonon couplings depend on the electron density of states, the phonon frequency and the linewidth of relevant modes. As the material evolves into the metastable state, the strong renormalization of both the electron and phonon spectra opens up the possibility to adjust the strength of the different coupling channels. These changes in coupling originate from exploiting the light-induced modifications of the electron and phonon systems, which are sensitively captured by our technique that maps the electron temperature and band structure. Our approach can be extended to investigate and control electron-boson interactions in other complex materials, and thus enable new routes for steering a quantum system towards a desired state using light.

**Acknowledgements:** We thank F. Carbone, S. Fahy, Y. Zhu and J. K. Freericks for useful discussions. We gratefully acknowledge support from the National Science Foundation through the JILA Physics Frontiers Center PHY-1125844, and a Gordon and Betty Moore Foundation EPiQS Award GBMF4538. Z. Tao also gratefully acknowledges the support from the National Natural Science Foundation of China (grant no. 11874121). F.C.K., P.M. and P.M.O. acknowledges support from the Swedish Research Council (VR), the K. and A. Wallenberg Foundation (grant No. 2015.0060) and the Swedish National Infrastructure for Computing (SNIC).

**Financial interests:** H.K. and M.M. have a financial interest in a laser company, KMLabs, that produces the lasers and HHG sources used in this work. H.K. is partially employed by KMLabs. The authors declare that they have no other competing or financial interests.

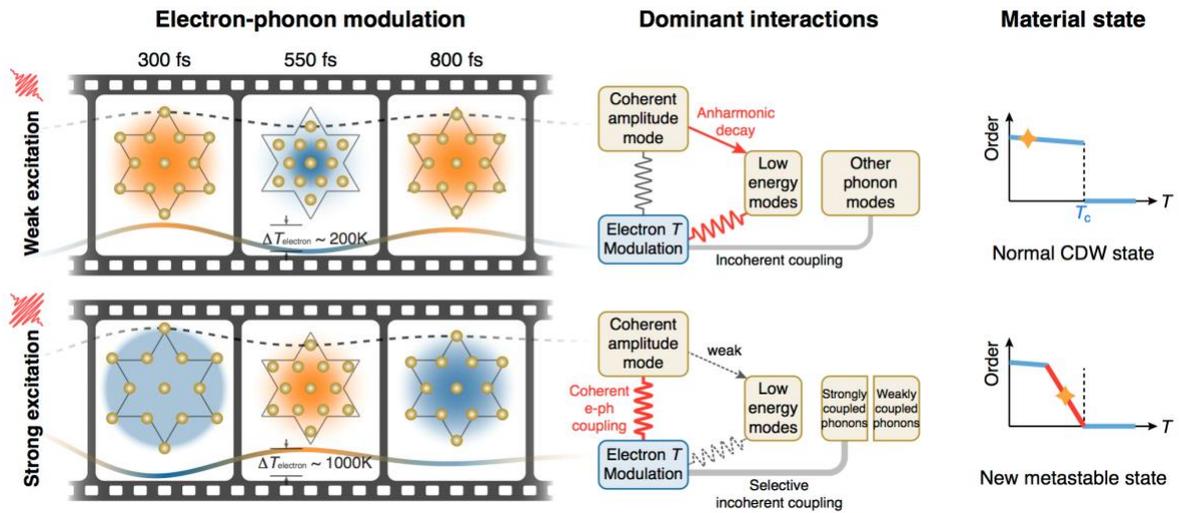

**Figure 1 | Schematic of the coherent electron-phonon modulation and dominant interactions.** The atomic displacement and charge density in a star-of-David at selected time delays illustrates how a modulation of the periodic lattice distortion (CDW amplitude mode) changes the electron temperature. The brown circles represent Ta atoms, the colour shading represents the charge density and the colour indicates the oscillatory part of the electron temperature, all amplitudes are exaggerated for better visualization. The electron temperature and the amplitude mode oscillate in phase at low fluences, while they oscillate in antiphase at the fluences higher than a critical point. This π phase shift is associated with a switching of dominant interaction from an anharmonic phonon-phonon coupling to a coherent electron-phonon coupling, as well as an ultrafast CDW transition to a new metastable state. *T* in the right panel represents the effective temperature.

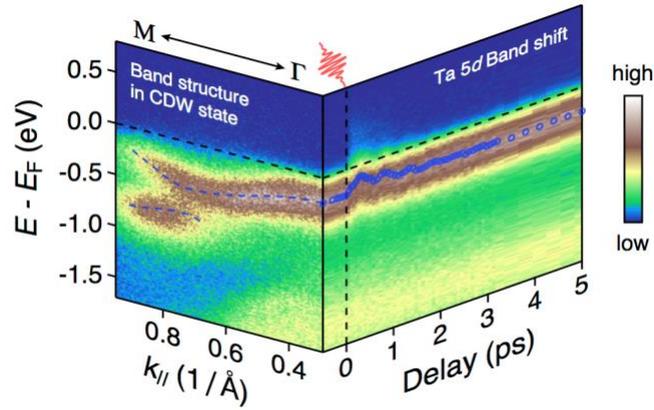

**Figure 2 | Coherent amplitude mode revealed by trARPES.** Experimental ARPES intensity plot along the Γ-M direction (left) and temporal evolution of the spectrum at the momentum $k_{//}$ around 0.3 Å$^{-1}$ after laser excitation with a fluence of 0.4 mJ/cm$^2$ (right). The red circles represent the extracted band positions, where a band oscillation that is coupled to the amplitude mode can be clearly observed.

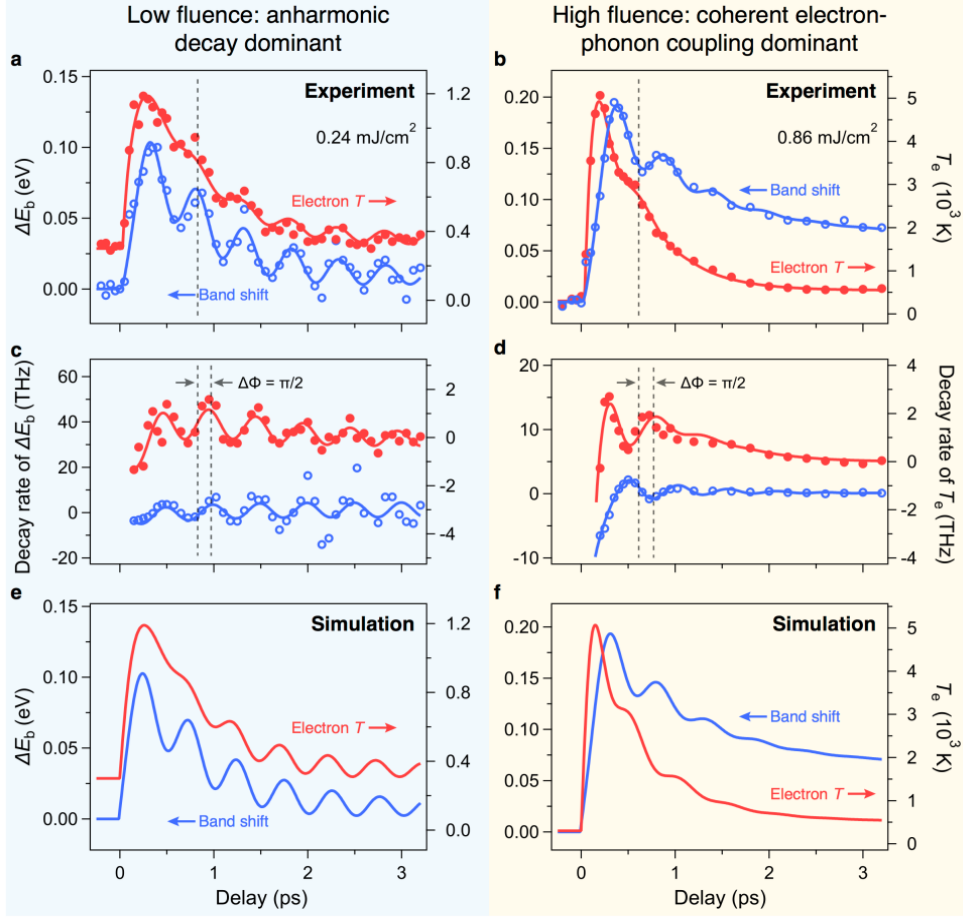

**Figure 3 | Coherent modulation of the electron temperature driven by the amplitude mode. a**, Simultaneously extracted binding energy shift of the Ta 5$d$ band (blue, left axis) and electron temperature (red, right axis), as a function of time delay after laser excitation with a fluence of 0.24 mJ/cm$^2$. The red and blue curves are fits to the data, see text. **b**, Same as **a** but with a higher fluence of 0.86 mJ/cm$^2$. **c** and **d**, Instantaneous decay rates determined by taking the logarithmic derivative of the data in **a** and **b** with respect to time delay, respectively. The oscillation in the electron temperature can be clearly observed at both laser fluences. It is coherently locked to the band oscillation, in phase at 0.24 mJ/cm$^2$ while in antiphase at 0.86 mJ/cm$^2$. **e** and **f**, Simulations of the electron temperature dynamics based on Eq. (1). **e**, At low fluences, the $T_e$ oscillation can be reproduced by assuming a dominant contribution from anharmonic decay of the coherent phonon mode, which is in phase with the band shift. **f**, At high fluences, the $T_e$ oscillation and π phase change can be reproduced by assuming a dominant contribution from coherent electron-phonon coupling.

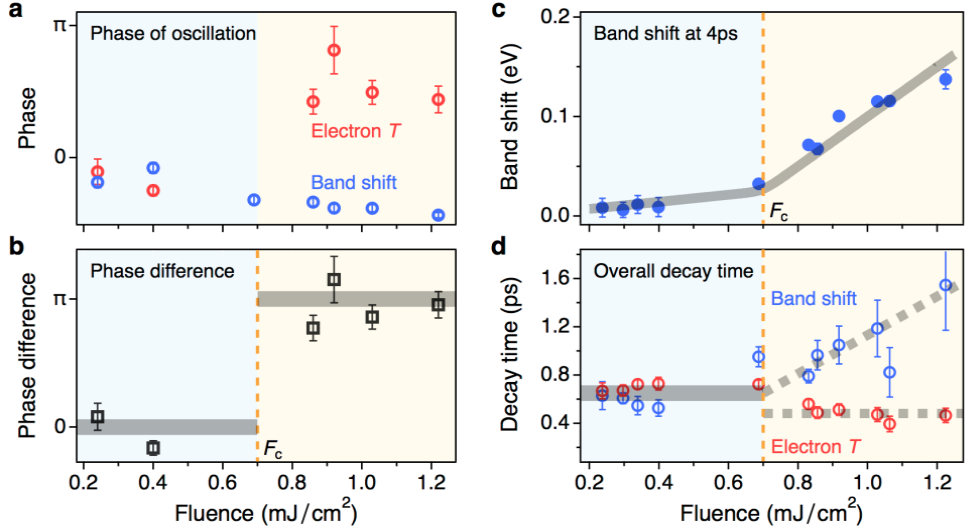

**Figure 4 | π phase change of the electron temperature modulation associated with the metastable state. a**, Fitted phases of the oscillations in the band shift (blue) and electron temperature (red) as a function of laser fluence. **b**, Phase difference between these two oscillations. It is around 0 (in phase) at low fluences and switches to π (in antiphase) at fluences higher than a critical fluence $F_c$. **c**, Band shift at 4 ps, and **d**, overall relaxation time constant of band shift and electron temperature, as a function of fluence. When $F > F_c$, the overall decay time of the electron temperature decreases, the decay of band shift deviates from that of electron temperature, and the material evolves into a metastable state, see text.